\documentclass{article}

\usepackage{microtype}
\usepackage{graphicx}
\usepackage{subfigure}
\usepackage{booktabs} 

\usepackage{hyperref}

\usepackage[accepted]{icml2019}

\icmltitlerunning{Detecting anthropogenic cloud perturbations with deep learning}

\begin{document}

\twocolumn[
\icmltitle{Detecting anthropogenic cloud perturbations with deep learning}



\icmlsetsymbol{equal}{*}

\begin{icmlauthorlist}
\icmlauthor{Duncan Watson-Parris}{equal,ox}
\icmlauthor{Samuel Sutherland}{equal,ox}
\icmlauthor{Matthew Christensen}{ox}
\icmlauthor{Anthony Caterini}{oxs}
\icmlauthor{Dino Sejdinovic}{oxs}
\icmlauthor{Philip Stier}{ox}
\end{icmlauthorlist}

\icmlaffiliation{ox}{Atmospheric, Oceanic and Planetary Physics, Department of Physics, University of Oxford, Oxford, UK}
\icmlaffiliation{oxs}{Department of Statistics, University of Oxford, UK}

\icmlcorrespondingauthor{Duncan Watson-Parris}{duncan.watson-parris@physics.ox.ac.uk}

\icmlkeywords{Deep Learning, ICML, Climate}

\vskip 0.3in
]



\printAffiliationsAndNotice{\icmlEqualContribution} 

\begin{abstract}

One of the most pressing questions in climate science is that of the effect of anthropogenic\footnote{A term commonly used in climate science to mean caused by human activity.} aerosol on the Earth's energy balance. Aerosols provide the `seeds' on which cloud droplets form, and changes in the amount of aerosol available to a cloud can change its brightness and other physical properties such as optical thickness and spatial extent. Clouds play a critical role in moderating global temperatures and small perturbations can lead to significant amounts of cooling or warming. Uncertainty in this effect is so large it is not currently known if it is negligible, or provides a large enough cooling to largely negate present-day warming by CO2. This work uses deep convolutional neural networks to look for two particular perturbations in clouds due to anthropogenic aerosol and assess their properties and prevalence, providing valuable insights into their climatic effects.
\end{abstract}

\section{Introduction}
\label{intro}

The planetary energy balance (between incoming and outgoing radiation), and hence global temperature, is very sensitive to the properties and distribution of clouds in the atmosphere. In turn, the properties of clouds depend on the availability of condensation nuclei in the form of aerosol \cite{Lohmann2016}. An increase in aerosol particles, due to human activity for example, generally enhances the number of small cloud droplets for a given amount of liquid water. Smaller droplets are more reflective and so increase the albedo of the cloud \cite{twomey_pollution_1974}. These effects are difficult to observe though because they are small compared to the other main drivers of cloud formation, such as moisture convergence and the availability of condensable water. These changes also affect the availability of aerosol, and clouds themselves change, remove, and even produce aerosol, creating complex feedbacks which are hard to disentangle.

One approach to tackle this problem is by looking for strong local aerosol perturbations in otherwise clean environments, ideally with slowly varying cloud properties. Ship tracks –- tracks of cloud made brighter than their surroundings by ship emissions, provide ideal cases \cite{Christensen:2011by}. Finding and analysing these tracks is, however, a laborious and time-consuming process, and hence no global long-term databases of these important phenomena exist. 

Another, even less well understood, process by which changes in aerosol amount can change cloud properties is the so called lifetime effect. Cloud droplets will start to precipitate once they reach a certain critical diameter, usually around 14$\mu$m. All other things being equal, a smaller distribution of cloud droplets can delay the onset of precipitation in a cloud, potentially enhancing its lifetime and fractional area \cite{ALBRECHT1227}. This acts to increase the albedo of the cloud further, potentially creating a large cooling effect (relative to the unperturbed cloud) \cite{ipcc_chap7}. 

Large continuous decks of stratocumulus (Sc) clouds occur in the cold upwelling regions of the major oceans and play a crucial role in the global energy balance because of their size and persistence. However, an important phenomena occurs where a large region of the cloud deck dissipates through the onset of precipitation and leaves open regions, so called Pockets of Open Cells (POCs) \cite{Stevens:2005cca}. It has been hypothesized that these occurrences could be affected by anthropogenic activity, specifically through aerosol perturbations delaying or even inhibiting their onset and causing a large cooling effect through the mechanisms described above \cite{acp-6-2503-2006}. Hence, POCs could have important implications for climate change. Again however, besides from a handfull of small-scale case studies, no database of POC occurrence exists. 

Here we present a method of automatically detecting ship tracks and POCs in 140 TB of high resolution satellite imagery using deep convolutional neural networks. This model is then run over many years of satellite data producing a unique database of observations that allows detailed studies of the probability of their occurrence in different regions and physical conditions. These implications will feed back into regional and global assessments of their effect on global cloud forcing.

The following section will outline the data used and methodology developed for the detection of these phenomena, before moving on to a presentation of the main findings so far in Section \ref{results}. We conclude with a discussion of the implications of our work and an outlook in Section \ref{conclusion}.

\section{Methodology}
\label{methods}

\subsection{Data}
\label{data}

All of the data used in the methods outlined below were obtained from the Moderate Resolution Imaging Spectrometer (MODIS) instruments on the NASA Aqua and Terra \cite{MOD02} satellites. The Level 1B data sets were used which provide calibrated and geolocated radiances at-aperture for all 36 MODIS spectral bands at 1km resolution.  Due to the different scales and properties of the two different phenomena, however, the data were prepared independently. The models used are also slightly different in each case, though in the future we hope to combine them into a single model.

For the ship track detection dataset, the original imager reflectances from Channels 1, 20 and 32 (corresponding to wavelengths of 645nm, 3.75$\mu$m and 12$\mu$m respectively) were combined into a three-channel false-color composite. This composite was designed to provide information in the visible (towards the middle of the solar spectrum), the near-infrared (which provides information about the cloud droplet size) and the infra-red (which allows discrimination of cloud liquid and ice). The original 1350x2030 pixel images were split into 12 440x440 pixel padded images to reduce the size of image used in training, while maintaining the full 1km resolution. The training data was provided in the form of 4,500 hand-logged tracks \cite{Segrin2007,Toll2017,Christensen2012} which were interpolated and converted into pixel masks for use in training the model. An example image and the corresponding hand-logged data is shown in Fig. \ref{fig:inference_examples}b.

For the POC detection dataset, the micro-physical properties of the cloud were less important and a true-color RGB composite was used from channels 1, 4 and 3 (corresponding to wavelengths of 645nm, 555nm and 469nm respectively). We use SatPy \cite{Raspaud2018} to prepare the composite images from the raw HDF4 datafiles. Due to the relatively large size of the features and to speed up training the images were rescaled from 1350x2030 pixels to 648x1296 and then split in two to create images 648x648 in size, which were then further rescaled to 224x224 to match ResNet-152. The training dataset was created by hand-logging 1029 images which contained 715 POCs (as shown in Fig. \ref{fig:inference_examples}a).

\subsection{Model}
\label{model}

Both of the models used are based on ResUnet architecture which has shown good performance in remote observation settings \cite{ResUNet}. 

\begin{figure}
    \centering
    (a)\includegraphics[width=0.15\textwidth]{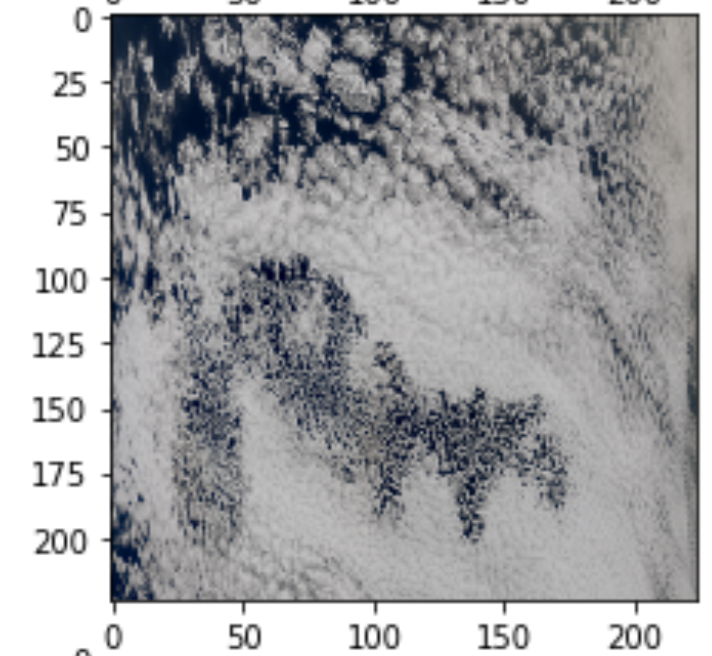}
    \includegraphics[width=0.15\textwidth]{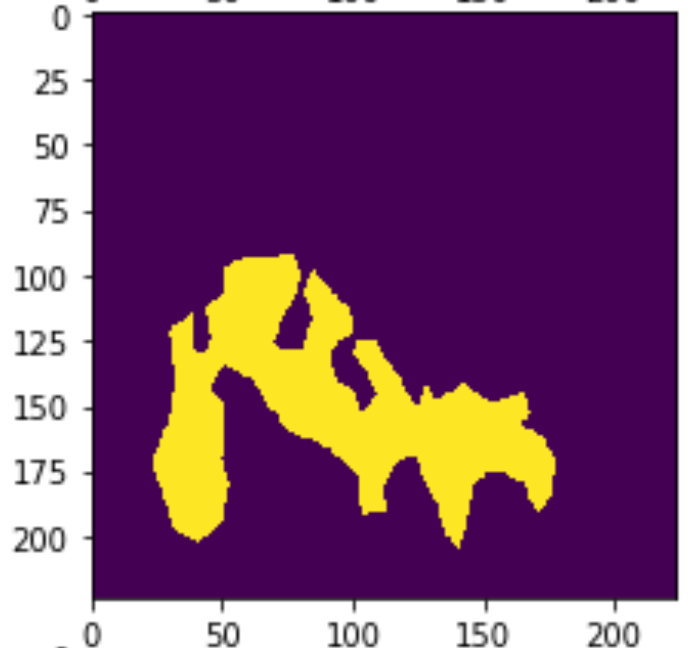}
    \includegraphics[width=0.15\textwidth]{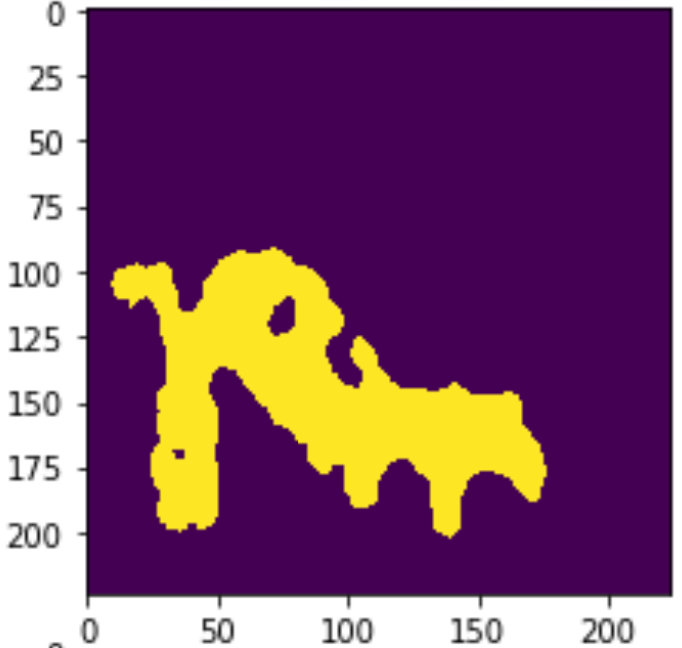}
    (b)\includegraphics[width=0.45\textwidth]{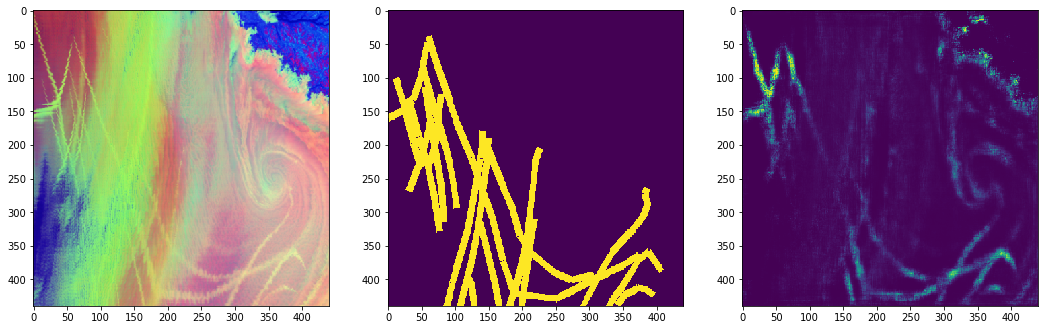}
    \caption{From left to right: An example input image, the hand logged validation mask, and  the inferred mask from the model for both the POC model (a) and the ship track model (b). Note that not all the ship tracks are included in the hand logged mask due to human error and/or the selection criteria used.}
    \label{fig:inference_examples}
\end{figure}

The ship-track detection model currently uses the ResUnet model with a binary cross-entropy loss function and train it over 100 epochs. We use the Adam optimization \cite{kingma2014adam} with a step-down in the learning rate from 1e-5 to a minimum of 5e-7 in fractional steps of 0.2 when the loss plateaus. While this model shows some qualitative skill, as shown in Fig. \ref{fig:inference_examples}b, at the time of writing training and development was still ongoing.

The POC detection model uses a modified ResNet-152 whose dense layers have been removed and replaced by three up-sampling blocks based on the second half of the ResUnet model used for the ship-track detection. The ResNet-152 portion of the model had been pre-trained on ImageNet, since this gives strong texture recognition, which is important in this case as the different cloud structures give very different textures. The upsampling blocks were trained using the Adam optimization algorithm and step-down learning rate from 0.001 with fractional steps of 0.2. The loss function used was the DICE coefficient, as it is robust to class imbalance and gave very good results in testing. The final masks were refined using a reduced ResUnet model that was trained in the same way as the ResNet-152 model. The model obtains a precision score of 0.73 on the test set, but visually performs extremely well. 8,491 POCs were found in the 25,582 files on which the model was run, creating the largest dataset of this phenomenon to date.

\section{POC Results and Analysis}
\label{results}

By applying the POC detection model to all of the MODIS images which intersect the main Sc regions off the coast of California, Peru and Namibia we are able to build a large database of the temporal and spatial distribution of POCs and their properties. Here we breifly summarise their main characteristics.

\begin{figure}
    \centering
    \includegraphics[width=0.45\textwidth]{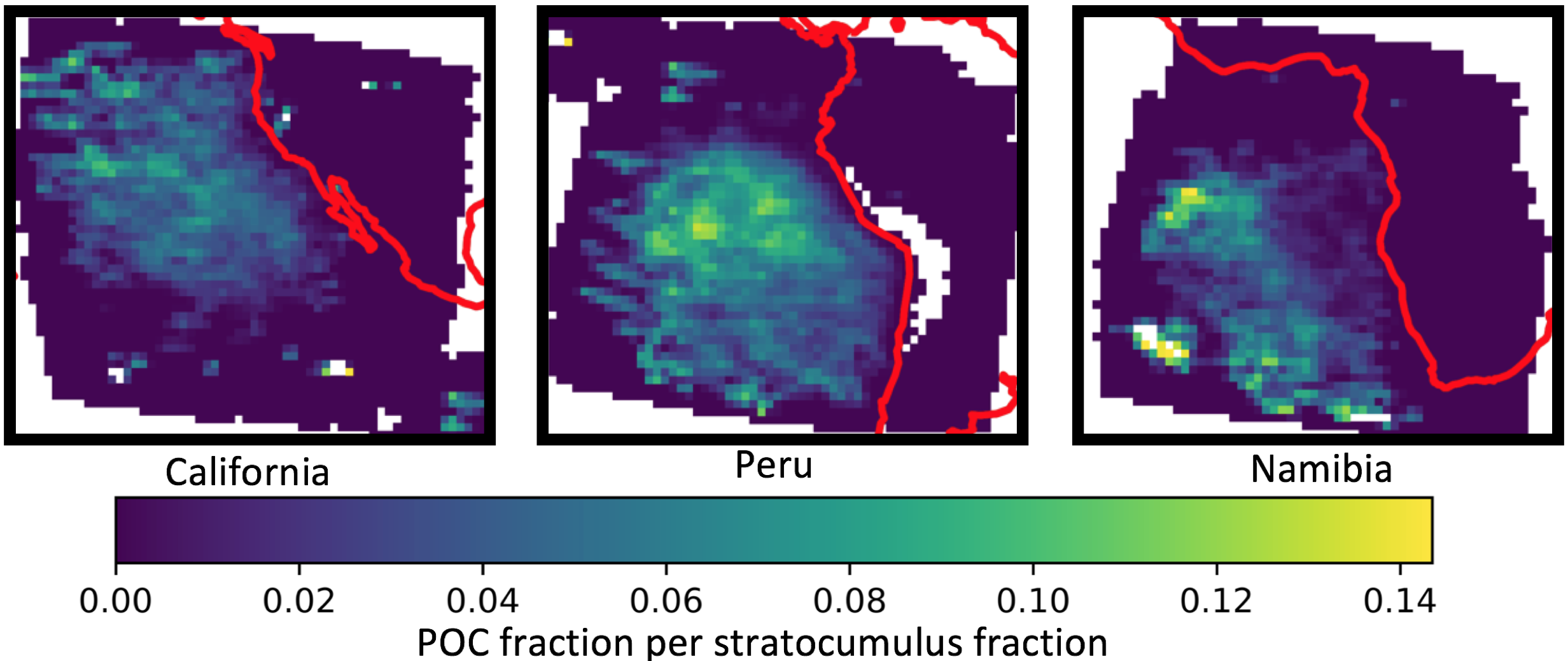}
    \caption{The density of POCs in each of the main Stratocumulus (Sc) regions is shown normalised by the number of images used for inference and the climatological fraction of Sc.}
    \label{fig:POC_density}
\end{figure}

Figure \ref{fig:POC_density} shows the density of POCs in each of the main Sc regions, normalized by the number of images used and the climatological fraction of Sc \cite{ISCCP1983}. Both the Californian and Namibian POCs show highest densities towards the edge of the cloud decks, whereas the Peruvian POCs are more evenly spread across the deck. The Peruvian Sc region also shows the highest number of POCs overall. The temporal distribution of POCs shows a strong seasonal cycle and peaks in the local winter, although POCs in the Californian Sc deck also have a peak in northern hemisphere summer (not shown).

\begin{figure}
    \centering
    \includegraphics[width=0.4\textwidth]{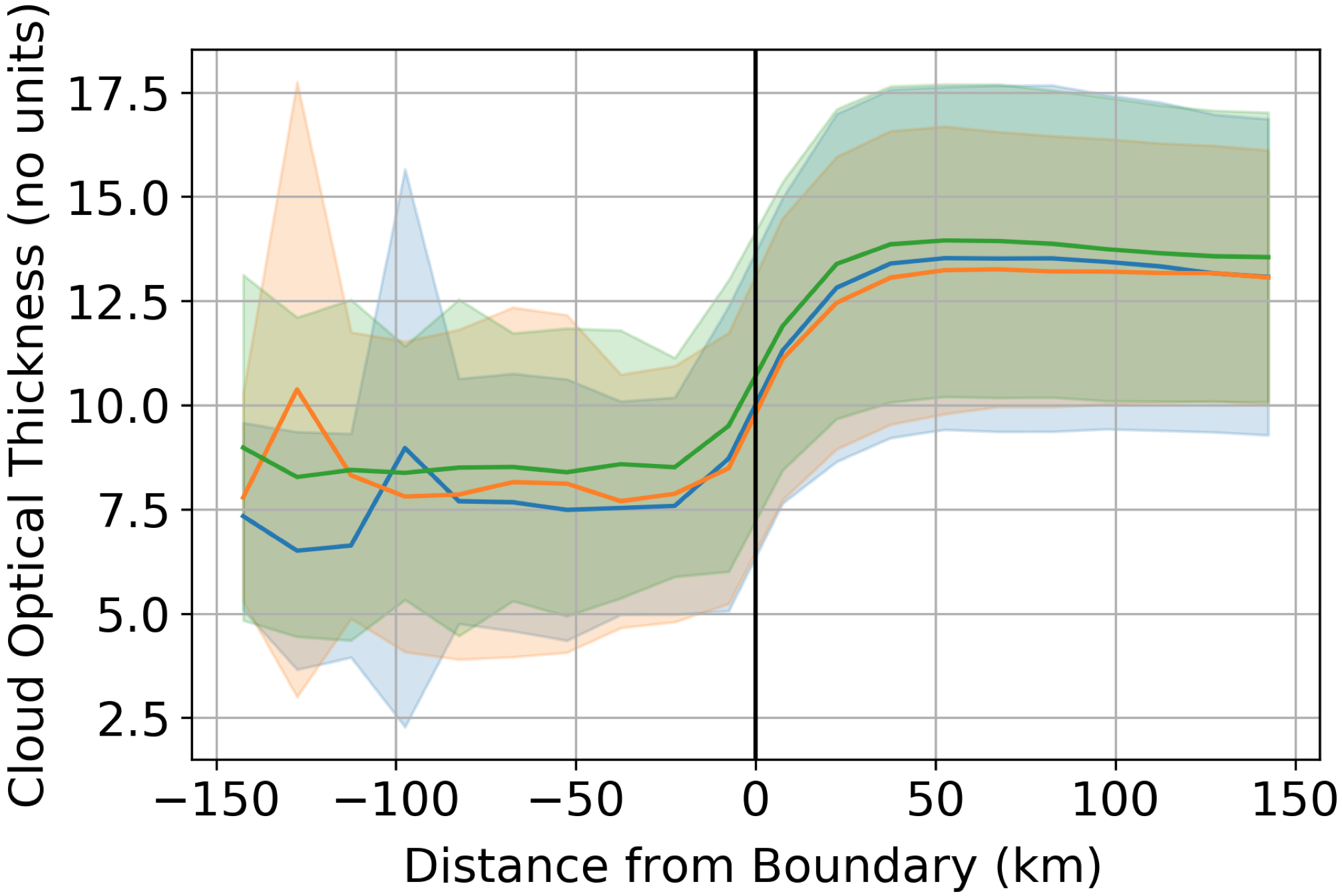}
    \caption{The mean cloud optical thickness as a function of distance from the POC boundary. Shading indicates the standard deviation in the mean over each distance bin. Each color represents a region.}
    \label{fig:POC_tau}
\end{figure}

A number of retrievals are performed on the raw MODIS spectral radiances to determine cloud properties at each pixel. By applying the inferred POC masks to the retrieved MODIS cloud properties \cite{MOD06_L2} we are able to build statistics about the POCs and their surrounding environment. Using OpenCV \cite{opencv_library} to extract regions of fixed distance from each POC we can plot the average properties as a function of distance from all of the detected POCs. Figure \ref{fig:POC_tau} shows the average retrieved cloud optical thickness for all of the POCs binned as a function of distance from the POC boundary. The increase in optical thickness inside the POC is clearly evident and due to the reduction in liquid water content through precipitation. The similarity between the three regions is striking and implies a common mechanism driving POC formation.

By combining the spatio-temporal distribution of POCs with their average optical depth and an assumed cloud droplet asymmetry parameter we are able to calculate the change in albedo due to POCs \cite{stephens1994remote}. Calculating a global mean value and multiplying by the incident solar flux, we find that POCs make a very small change to the amount of energy reflected by the Sc, only 0.02 Wm$^{-2}$. Even if anthropogenic activity suppressed POC formation completely the climate effect would be negligible compared to a present day CO2 forcing of 1.7 Wm$^{-2}$.

\section{Conclusions and discussion}
\label{conclusion}

Automatically detecting, and even classifying, clouds in satellite data is easy and done routinely, however the detection of \emph{perturbations} in those clouds is more challenging. In this work we have demonstrated the first application of deep convolutional neural networks in the detection of ship-tracks and POCs. The detection of ship tracks provides positive cases of the direct perturbation of clouds by anthropogenic aerosol, while the detection of POCs provides clues about the adjustments of clouds to these perturbations. 

By running inference over all images from the MODIS satellite intersecting the main Sc regions we have created a database of the properties of 1000s of POCs and their spatial and temporal characteristics. This has enabled a global, long term analysis of these phenomena, which will provide the keys to unlocking new understanding of the conditions under which aerosols influence cloud physics and ultimately the climate.

While the ship track model is currently still under development, early results provide encouraging evidence that the model has some skill and could be improved further using some of the techniques developed for POC detection. 

Looking forward, we hope to develop a single model which is able to detect both these, and other cloud perturbations in order to produce a global, open, database spanning more than a decade of satellite observations. This will prove invaluable in our efforts to better constrain the effects of aerosol on clouds and their overall contribution to anthropogenic climate change.

\section*{Acknowledgements}

We gratefully acknowledge the support of Amazon Web Services through an AWS Machine Learning Research Award. We also acknowledge the support of NVIDIA Corporation with the donation of a Titan Xp GPU used for this research. DWP and PS acknowledge funding from the Natural Environment Research Council project NE/L01355X/1 (CLARIFY). PS and MC acknowledge funding from the European Research Council project RECAP under the European Union's Horizon 2020 research and innovation programme with grant agreement 724602.

\subsection*{Software and Data}

The MODIS data used were the 021km and 06 products available through https://modis.gsfc.nasa.gov/data/. The models were built using Keras \cite{chollet2015keras} with the Tensorflow engine \cite{tensorflow2015-whitepaper}. The models and training data will be made freely available on publication.

\bibliography{paper}
\bibliographystyle{icml2019}

\end{document}